\documentstyle[epsfig]{new}
\begin{document}
\def\ltap{\ \raisebox{-.4ex}{\rlap{$\sim$}} \raisebox{.4ex}{$<$}\ }
\def\gtap{\ \raisebox{-.4ex}{\rlap{$\sim$}} \raisebox{.4ex}{$>$}\ }

\title{ %
The Oscillation Length Resonance in the Transitions
of Solar and Atmospheric Neutrinos Crossing the Earth Core} 

\author{Serguey PETCOV
\footnote{Talk given at the Satellite Symposium
           ``New Era in Neutrino Physics'',
           June 11 - 12, 1998, Tokyo Metropolitan University, Tokyo, Japan
           (to be published in the Proceedings of the Symposium).}
{\it Scuola Internazionale Superiore di Studi Avanzati, \\
 Istituto Nazionale di Fizica Nucleare, Sezione di Trieste,\\
 I-34013 Trieste, Italy, and
 Bulgarian Academy of Sciences, \\
  1784 Sofia, Bulgaria, \\
petcov@susy.sissa.it}}

\maketitle

\section*{Abstract}

  The characteristic  
features of the neutrino 
oscillation length resonance, taking place 
in the $\nu_2 \rightarrow \nu_{e}$ and
$\nu_{\mu} \rightarrow \nu_{e}$ 
($\nu_e \rightarrow \nu_{\mu (\tau)}$)
transitions in the Earth 
of the solar and atmospheric neutrinos  
crossing the Earth core,
are reviewed. The resonance enhances dramatically these
transitions at small mixing angles 
but differs from the MSW one. 
It can be responsible, e.g., for the small excess of 
$e-$like events at Zenith angles $\theta_{z} \sim (130^{\circ} - 180^{\circ})$
and can produce at least part of the Zenith angle
dependence of the rate of $\mu-$like events,
exhibited by the Super-Kamiokande atmospheric 
neutrino data.

\section{Introduction}
  The $\nu_2 \rightarrow \nu_{e}$ and 
$\nu_{\mu} \rightarrow \nu_{e}$
($\nu_e \rightarrow \nu_{\mu(\tau)}$) transitions of 
the solar and atmospheric neutrinos in the Earth,
caused by neutrino mixing 
in vacuum \footnote{The $\nu_2 \rightarrow \nu_{e}$
transition probability accounts,
as is well-known, for the Earth effect 
in the solar neutrino MSW survival probability
in the case of the  
two-neutrino  $\nu_e \rightarrow \nu_{\mu(\tau)}$ and
$\nu_e \rightarrow \nu_{s}$ transition 
solutions of the solar neutrino problem, 
$\nu_{s}$ being a sterile neutrino.},
are enhanced by a new type of resonance
which differs from the MSW one and takes place 
when the neutrinos cross the 
Earth core [1]. 
The enhancement is particularly strong 
in the case of small  
mixing [1]. The conditions for existence of the new resonance
include specific constraints on the neutrino
oscillation length (and mixing angles) in the 
Earth mantle and core.
When satisfied, these constraints ensure that certain  
oscillating factors in the
probabilities of the $\nu_2 \rightarrow \nu_{e}$ and 
$\nu_{\mu } \rightarrow \nu_{e}$
($\nu_e \rightarrow \nu_{\mu(\tau)}$) transitions,
$P_{e2}$ and $P(\nu_{\mu (e)} \rightarrow \nu_{e (\mu;\tau)})$,
have maximal values and that this 
produces a resonance maximum in $P_{e2}$ and 
$P(\nu_{\mu (e)} \rightarrow \nu_{e (\mu;\tau)})$.
Correspondingly, the term ``neutrino oscillation 
length resonance'' (NOLR) 
or simply ``oscillation length resonance'' 
was used in [1] 
to denote the resonance of interest. 
The NOLR is in many respects similar, 
e.g., to the resonance transforming a  
circularly polarized LH photon into a RH one 
when the photon traverses three layers of optically active 
medium such that the optical activity and  
the widths of the first and third layers are identical 
but differ from those of the second layer
\footnote{There exists also a 
beautiful analogy between the NOLR 
and the electron 
spin-flip resonance realized  
in a specific configuration of 
magnetic fields 
(see [1] for a more detailed discussion).}.

  The present article contains a brief review 
of the specific features of the 
oscillation length resonance 
in the simplest case of 
two-neutrino transitions of the Earth core crossing
solar and atmospheric neutrinos.
 We discuss also the 
implications of the 
NOLR enhancement of these transitions [1,2,3,4] 
for the interpretation of the results of the 
Super-Kamiokande experiment [5,6]. 

\section{The Neutrino Oscillation Length Resonance (NOLR) in the
Transitions of Solar and Atmospheric Neutrinos Traversing the Earth}

   According to the existing Earth models [7], 
the matter density and 
the chemical composition change 
mildly within the two
main density structures in the Earth - the mantle and the core.
As a consequence, all the interesting features of the
solar and atmospheric neutrino transitions
in the Earth, including those related to
the neutrino oscillation length resonance,
can be understood quantitatively
assuming that the neutrinos 
crossing the mantle and the core traverse effectively
two layers with constant but different densities,
$\bar{\rho}_{man}$ and $\bar{\rho}_{c}$,
and  chemical composition or electron fraction numbers,
$Y_e^{man}$ and $Y_e^{c}$ [1,3,8,9].
The neutrino transitions of interest 
in the two-layer model
of the Earth density distribution 
\footnote{The densities $\bar{\rho}_{man,c}$
should be considered as mean
densities along the neutrino trajectories.
In the Earth models [7] 
one has: $\bar{\rho}_{man} \cong (4 - 5)~ {\rm g/cm^3}$ and
$\bar{\rho}_{c} \cong (11 - 12)~ {\rm g/cm^3}$.
For $Y_e$ one can use the standard
values [7,10] 
(see also [4]) 
$Y_e^{man} = 0.49$ and $Y_e^{c} = 0.467$.} 
result from two-neutrino oscillations
taking place i) first in the mantle over a distance $X'$
with a mixing angle $\theta'_{m}$ and oscillation length
$L_{man}$, ii) then in the core over a
distance $X''$ with different mixing angle $\theta''_{m}$
and oscillation length $L_{c}$, and iii) again
in the mantle over a distance $X'$
with $\theta'_{m}$ and $L_{man}$.
Because $\bar{\rho}_{man,c} \neq 0$, 
$\bar{\rho}_{man} \neq \bar{\rho}_{c}$ and
$Y_e^{man} \sim Y_e^{c}$, and due to the matter effect, 
we have for given neutrino
energy $E$, mass squared difference $\Delta m^2 > 0$ and
vacuum mixing angle $\theta$, $\cos 2\theta > 0$ 
(see, e.g., [11]): 
$\theta'_{m},\theta''_{m} \neq \theta$,    
$\theta'_{m} \neq \theta''_{m}$, $L_{man}\neq L_{c}$
and $L_{man,c}\neq L^{vac}$, $L^{vac} = 4\pi E/\Delta m^2$ being 
the neutrino oscillation length in vacuum. 
For fixed $X'$ and $X''$
the oscillation length
resonance occurs [1] 
if i) the relative phases acquired by the
energy eigenstate neutrinos
in the mantle and in the core,
$\Delta E'X'~ = 2\pi X'/L_{man}$ and $\Delta E''X'' = 2\pi X'' /L_{c}$,
are correlated,
$$\Delta E'X'~ = \pi (2k + 1),~~~\Delta E''X'' = \pi (2k' + 1), 
~~~k, k' = 0,1,2,...,~~~\eqno(1)$$
\noindent and if ii) the inequalities
$$\sin^2(2\theta''_{m} - 4\theta'_{m} + \theta) - \sin^2\theta > 0,
~~\eqno(2a)$$
$$\sin^2(2\theta''_{m} - 4\theta'_{m}  + \theta)\sin(2\theta''_{m} -
2\theta'_{m})\sin(2\theta'_{m} - \theta)
\cos(2\theta''_{m} - 4\theta'_{m} + \theta)~~$$
$$~+ \frac{1}{4}\sin^2\theta \sin(4\theta''_{m} - 8\theta'_{m} + 2\theta)
~\sin(4\theta''_{m} - 4\theta'_{m}) < 0.~~\eqno(2b)$$
\noindent are fulfilled. Conditions (1) are universal, i.e., they are valid
for $P_{e2}$, $P(\nu_{\mu} \rightarrow \nu_{e}) = 
P(\nu_{e} \rightarrow \nu_{\mu (\tau)})$, etc. while (2a) and (2b)
correspond only to $P_{e2}$ [1].  
When equalities (1) hold, (2a) and (2b) ensure that $P_{e2}$ has a maximum. 
In the region of the NOLR maximum, e.g., for
$\Delta E'X' \cong \pi(2k + 1)$,
$P_{e2}$ is given 
by [1]: 
$$P_{e2} \cong \sin^2\theta~+~
{1\over {4}} \left [1 - \cos \Delta E''X'' \right ] 
\left [ \sin^2(2\theta''_{m} - 4\theta'_{m}  + \theta) -
\sin^2\theta \right ].~\eqno(3)$$
\noindent At the  NOLR maximum we have [1] (Fig. 1):
$$P^{max}_{e2} = \sin^2(2\theta''_{m} - 4\theta'_{m}  + \theta).~\eqno(4)$$
The analogs of eqs. (2) - (4) 
for $P(\nu_{\mu (e)} \rightarrow \nu_{e (\mu;\tau)})$ 
can be obtained 
by formally setting $\theta = 0$ 
while keeping $\theta_{m}',\theta_{m}'' \neq 0$ 
in eqs. (2) - (4):
thus, (2a) becomes $\sin^2(2\theta''_{m} - 4\theta'_{m}) > 0$ 
and is always fulfilled, while
(2b) transforms into [1] 
$$\cos(2\theta''_{m} - 4\theta'_{m}) < 0.~~\eqno(5)$$
   
   Note that  since
$X'$ and $X''$ are fixed
for a given neutrino trajectory, 
which is specified by its Nadir angle $h$, 
the NOLR conditions (1) are constraints
on the oscillation lengths  $L_{man}$ and  $L_{c}$.
At $\sin^22\theta \ltap 0.02$
one of the two requirements in (1) 
is equivalent for $k = k' = 0$  
to the physical condition [1] 
$$ \pi~({1\over {X'}} + {1\over {X''}})
\cong \sqrt{2} G_F {1\over {m_{N}}} (Y_e^{c}\bar{\rho}_{c} -
Y_e^{man}\bar{\rho}_{man}).$$ 

  Remarkably enough,
for the $\nu_2 \rightarrow \nu_{e}$ and 
$\nu_{\mu (e)} \rightarrow \nu_{e (\mu;\tau)}$ transitions 
of the solar and atmospheric neutrinos in the Earth, 
the NOLR conditions (1)  with $k=k'=0$ 
are approximately fulfilled at small 
$\sin^22\theta$ 
in the regions where (2a) - (2b) or (5) hold [1]. 
The associated NOLR maxima in  
$P_{e2}$ and $P(\nu_{\mu (e)} \rightarrow \nu_{e (\mu;\tau)})$ 
are absolute maxima and dominate in 
$P_{e2}$ and $P(\nu_{\mu (e)} \rightarrow \nu_{e (\mu;\tau)})$
[1,3] (Fig. 2):
the values of the probabilities at these maxima 
in the case of two-neutrino mixing
are considerably larger -
by a factor of $\sim (2.5 - 4.0)$ ($\sim (3.0 - 7.0)$), 
than the values of $P_{e2}$ and 
$P(\nu_{\mu (e)} \rightarrow \nu_{e (\mu;\tau)})$
at the local maxima 
\begin{figure}[t]
\vspace{-12pc}         
\centerline{ \epsfig{file=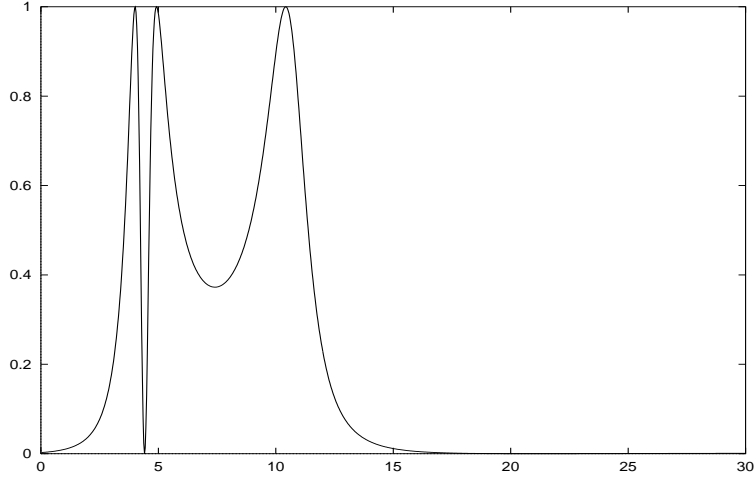,height=12cm,width=10cm,clip=,angle=0}}
\vspace*{-2.5pc}   
\caption{The dependence of the function
$\sin^2(2\theta''_{m} - 4\theta'_{m} + \theta)$, eq. (4), 
on $\rho_r 
= 1.32\times 10^{7}~\Delta m^2 [eV^2]~cos2\theta/E[MeV]~{\rm g/cm^3} \approx 
\rho^{res}_{man}$, $\rho^{res}_{man}$ being the 
resonance density in the 
Earth mantle [1], for $\sin^22\theta = 0.01$ (from [3]).} 
\end{figure}
associated with the MSW effect 
in the Earth core (mantle).
The magnitude of the NOLR 
enhancement depends on the neutrino trajectory
through the Earth core: the enhancement is maximal for
the center-crossing neutrinos [1,3]. 
Even at small mixing angles 
the resonance is relatively wide both 
in the neutrino energy 
and the Nadir angle [1,3] (Fig. 2).
The NOLR (relative) energy width is 
given for $\sin^22\theta < 0.05$ approximately by 
$${\Delta E \over {E_{max}}} \cong \left ( 2(1 +  
{\sqrt{2}\over{\pi}} {1\over {m_{N}}} G_F  
Y_e^{man}\bar{\rho}_{man} X') \right )^{-1}. ~\eqno(6)$$
\noindent It practically does not depend on $\sin^22\theta$
and varies slowly with the change of $h$: for
$h \ltap 25^{\circ}$ one has  
\footnote{For further details and 
a study of the NOLR effects 
in the transitions $\nu_{2} \rightarrow \nu_{e}$ 
and $\nu_e \rightarrow \nu_{s}$ ($\nu_e - \nu_{s}$ mixing),
and in $\bar{\nu}_{\mu} \rightarrow \bar{\nu}_{s}$
(or $\nu_{\mu} \rightarrow \nu_{s}$) 
at small mixing angles, see [1,3].}
$\Delta E/E_{max} \cong 0.30$. 

  Let us note that the dominating maximum 
due to the NOLR can
{\it now} be recognized in certain 
plots of the probabilities $P_{e2}$ and/or
$P(\nu_{\mu (e)} \rightarrow \nu_{e (\mu;\tau)})$, 
shown in some of the articles 
discussing the Earth effect in the 
transitions of solar and atmospheric neutrinos
(see, e.g., [12,13]). This maximum was 
invariably interpreted to be due to the MSW effect 
in the Earth core 
\footnote{Let us add also that 
the analysis performed in [1] 
and summarized above 
differs substantially from 
the studies [14] of the 
resonance enhancement of the 
$\nu_e \rightarrow \nu_{\mu (\tau)}$ transitions 
in matter varying 
{\it periodically} along the neutrino path 
(parametric resonance). 
The density change along the path of 
a neutrino crossing the Earth core is not periodic
(even in the two-layer model it falls short of 
making one and a half periods) and 
the results obtained in [14] 
do not apply to the
case of transitions of the 
Earth core crossing neutrinos, 
studied in [1].}
before the study [1].

   Finally, in [15] the $\nu_{\mu} \rightarrow \nu_{s}$
transitions in the Earth were considered
for $\sin^22\theta \cong 1$. It was noticed that
in the region where
$\sqrt{2} G_F \bar{N}_n^{man,c} 
\gg \Delta m^2/E$, $\bar{N}_n^{man,c}$ being the neutron number density,
a new maximum in $P(\nu_{\mu} \rightarrow \nu_{s})$ appears when
$\sqrt{2} G_F \bar{N}_n^{man(c)} 
X'^{('')} \cong 2\pi$,
found to hold at $h \sim 28^{0}$. The height of the maximum
is comparable to the heights of the other ``ordinary'' maxima
present in $P(\nu_{\mu} \rightarrow \nu_{s})$
for $\sin^22\theta \cong 1$. It is 
stated in [15] that the
effect does not take place in the
$\nu_{\mu (e)} \rightarrow \nu_{e (\mu;\tau)}$ transitions and
the case of solar neutrinos is not discussed. 
\vspace*{-2pt}
\section{The Super-Kamiokande Atmospheric Neutrino Data and
 the Neutrino Oscillation Length Resonance}
\vspace*{-4pt}
  The implications of the NOLR 
enhancement of the probability $P_{e2}$ for 
\begin{figure}[t]
\vspace*{-2pc}
\centerline{ \epsfig{file=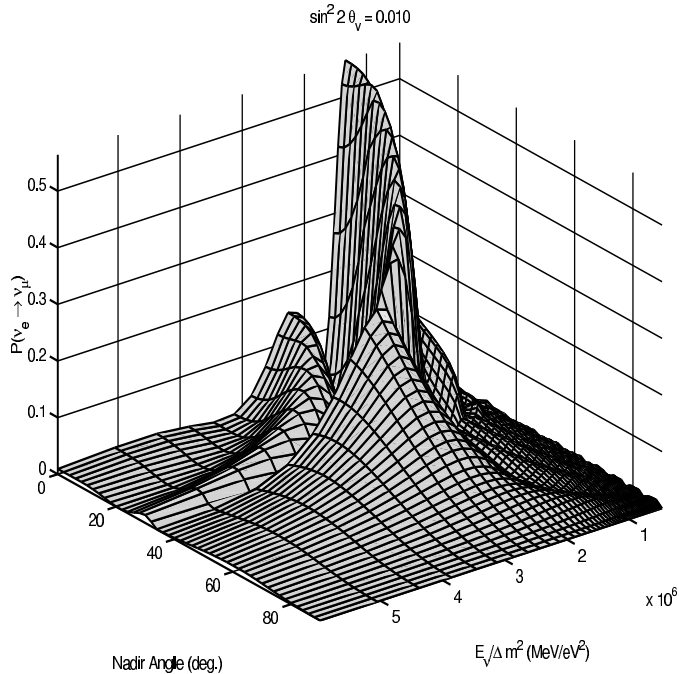,height=9cm,width=9cm}}
   \caption{The probability $P(\nu_{e (\mu)} \rightarrow \nu_{\mu;\tau (e)})$ 
as a function of $h$ and $E/\Delta m^2$ 
for $\sin^22\theta = 0.01$ [3]. The NOLR absolute maximum 
for $h \approx (0^0 -  28^0)$ is clearly seen 
at $E/\Delta m^2 \approx (1.3 - 1.6)\times 10^{6}~
{\rm MeV/eV^2}$. The local 
maximum at $E/\Delta m^2 \sim 2.5 
\times 10^{6}~{\rm MeV/eV^2}$ is due to the MSW effect in the Earth mantle.} 
\end{figure}
the tests of the MSW  
$\nu_e \rightarrow \nu_{\mu (\tau)}$ 
and $\nu_e \rightarrow \nu_{s}$  
solutions of the solar neutrino problem 
are discussed in detail in refs. [1,2,3,4]. 
In the case of the small mixing angle (nonadiabatic)
$\nu_e \rightarrow \nu_{\mu (\tau)}$ 
solution, the NOLR produces [4]
a much bigger - by a factor of $\sim 6$, 
day-night  
asymmetry in the 
Super-Kamiokande sample of solar neutrino events, 
whose night fraction is due to the {\it core-crossing} 
neutrinos, in comparison with
the asymmetry determined by using the 
{\it whole night} event sample.
The Super-Kamiokande collaboration
has successfully exploited 
this result [3] in the analysis of their solar 
neutrino data in terms of the MSW effect [5]. 

   The NOLR enhancement 
of the $\nu_{\mu (e)} \rightarrow \nu_{e (\mu;\tau)}$ 
transitions of the atmospheric neutrinos
crossing the Earth  
has important implications  
for the interpretation
of the Super-Kamiokande 
atmospheric neutrino data as well [1].  
The Zenith angle dependence 
of the rates of sub-GeV and multi-GeV 
$\mu-$like events, observed in 
the Super-Kamiokande experiment, 
provides a strong evidence 
for oscillations
of the atmospheric $\nu_{\mu}$ 
($\bar{\nu}_{\mu}$) [6]. 
The data is best described   
in the case of two-neutrino mixing
by $\nu_{\mu}(\bar{\nu}_{\mu}) 
\leftrightarrow \nu_{\tau}(\bar{\nu}_{\tau})$ 
vacuum oscillations with parameters 
$\Delta m^2 \cong (0.5 - 6.0)\times 10^{-3}~{\rm eV^2}$
and $\sin^22\theta \cong (0.8 - 1.0)$ [6].

  It is quite remarkable that for values of 
$\Delta m^2 \sim (0.5 - 6.0)\times 10^{-3}~{\rm eV^2}$
suggested by the Super-Kamiokande data, 
the NOLR strongly enhances the $\nu_{\mu} \rightarrow \nu_{e}$ (and 
$\nu_{e} \rightarrow \nu_{\mu (\tau)}$)
transitions of the atmospheric neutrinos crossing the Earth core,
making the transitions detectable even at small mixing angles [1].
Indeed, for, e.g., $\sin^22\theta = 0.01$,
$\Delta m^2 = 0.5;~1.0;~5.0\times 10^{-3}~{\rm eV^2}$, 
and $h = 0^{0}$ (Earth center crossing), 
the absolute maximum in 
$P(\nu_{\mu (e)} \rightarrow \nu_{e (\mu;\tau)})$ 
due to the NOLR 
takes place at $E \cong 0.75;~1.50;~7.5~$GeV.
The $\nu_e$ and $\nu_{\mu}$ with such energies 
contribute either to the sub-GeV or to the 
multi-GeV $e-$like and $\mu-$like Super-Kamiokande
event samples [6]. At small mixing angles
the NOLR enhancement holds 
practically for all neutrino 
trajectories through the Earth core [1,3]. 
Accordingly, it was suggested in [1,2] (see also [3])
that, depending on the value of $\Delta m^2$,
the small excess
of e-like events in the region $-1 \leq \cos\theta_{z}\leq -0.6$,
$\theta_{z} = \pi - h$ being the Zenith angle, either
in the sub-GeV or in the
multi-GeV sample, 
observed  (in both samples) in the
Super-Kamiokande experiment [6], 
is due to $\nu_{\mu} \rightarrow \nu_{e}$
small mixing transitions, $\sin^22\theta_{e\mu} \cong (0.01 - 0.10)$, 
strongly enhanced by the 
NOLR as neutrinos cross the Earth core on the  way
to the detector. 
The same NOLR enhanced transitions 
should produce at least part of the 
zenith angle dependence, 
exhibited  by the Super-Kamiokande 
sub-GeV and multi-GeV $\mu-$like data [2,3]. 

    The transitions of interest arise naturally, e.g., 
in a three-neutrino mixing scheme, in which 
the small mixing angle MSW 
$\nu_{e} \rightarrow \nu_{\mu}$ transitions 
with $\Delta m^2_{21} \sim (4 - 8)\times 10^{-6}~{\rm eV^2}$,
or large mixing angle 
$\nu_{e} \leftrightarrow \nu_{\mu}$ oscillations
with $\Delta m^2_{21} \sim 10^{-10}~{\rm eV^2}$,
provide the solution of the solar neutrino
problem and the atmospheric neutrino anomaly is
due to $\nu_{\mu} \leftrightarrow \nu_{\tau}$
oscillations with 
$\Delta m^2_{31} \sim (0.5 - 6.0)
\times 10^{-3}~{\rm eV^2}$ [1]. 
For $\Delta m^2_{31} \gg \Delta m^2_{21}$ the  
three-neutrino $\nu_{\mu} \rightarrow \nu_{e}$ 
and $\nu_{e} \rightarrow \nu_{\mu (\tau)}$
transition probabilities reduce to a two-neutrino
transition probability equal to $P(\nu_{e} \rightarrow \nu_{\tau})$ [16],
with $\Delta m^2_{31}$  and 
$\sin^22\theta_{13} =
4|U_{e3}|^2(1 - |U_{e3}|^2)$ playing 
the role of the 
two-neutrino oscillation parameters
\footnote{Analytic expression for the probability
$P(\nu_{e} \rightarrow \nu_{\tau})$
can be found in [1] 
(see also eq. (3)).},
$\theta_{13} \equiv \theta_{e\mu}$,
where $U_{e3}$ is the $e - \nu_3$ element of the lepton
mixing matrix $U$. The 
data [6,17] imply: 
$\sin^22\theta_{13} \ltap 0.20$
\footnote{Thus, searching for the 
$\nu_{\mu (e)} \rightarrow \nu_{e (\mu; \tau)}$
transitions of the atmospheric neutrinos,
amplified by the NOLR, 
can provide also unique information 
about the magnitude of 
$U_{e3}$ [18].}.
The fluxes of atmospheric $\nu_{e,\mu}$ with energy $E$,
crossing the Earth along a trajectory with Zenith angle 
$\theta_{z}$ before reaching the detector, $\Phi_{\nu_{e,\mu}}$,
are given in 
this scheme by [2,3,16]: 
$$\Phi_{\nu_e}  
\cong \Phi^{0}_{\nu_e} \left ( 1 + 
  [s^2_{23}~r(E,\theta_{z}) - 1]~ 
P(\nu_{e} \rightarrow \nu_{\tau}) \right ),~\eqno(7)$$ 
$$\Phi_{\nu_{\mu}} \cong \Phi^{0}_{\nu_{\mu}}( 1 + 
 s^4_{23}~ [(s^2_{23}~r(E,\theta_{z}))^{-1} - 1]~ 
P(\nu_{e} \rightarrow \nu_{\tau})~~~~~~~~~~~~~~~~~~~~~~~~~~~~~~$$ 
$$~~~~~~~~~~~~~~~~~~~~~~~~~-  2c^2_{23}s^2_{23}~[ 1 - 
Re~( e^{-i\kappa} 
A(\nu_{\tau} \rightarrow \nu_{\tau}))] ),~ \eqno(8)$$

\noindent where 
$\Phi^{0}_{\nu_{e(\mu)}} = \Phi^{0}_{\nu_{e(\mu)}}(E,\theta_{z})$ is the 
$\nu_{e(\mu)}$ flux in the absence of neutrino transitions/oscillations,
$s^2_{23} \equiv |U_{\mu 3}|^2/(1 - |U_{e3}|^2) \leq 1$,
$U_{\mu 3}$ being the $\mu - \nu_3$ element of $U$, 
$c^2_{23} = 1 - s^2_{23}$, and
$r(E,\theta_{z}) \equiv \Phi^{0}_{\nu_{\mu}}/ 
\Phi^{0}_{\nu_{e}}$.
The two-neutrino transition probability amplitude 
$A(\nu_{\tau} \rightarrow \nu_{\tau})$ is given in the two-layer
model of the Earth density distribution
by eq. (1) in ref. [1] in which first $\theta$ is formally set to 
$\pi/2$ while $\theta'_m$ and $\theta''_m$ are kept 
arbitrary, and after that $\sin\theta'_m$ ($\sin \theta''_m$) and 
$\cos\theta'_m$ ($\cos\theta''_m$) are interchanged. The phase
$\kappa$ has the form [2,3,16]:
$$\kappa \cong {1\over {2}} [ {\Delta m^2_{31}\over{2E}}~X + 
\sqrt{2} G_F {1\over{m_N}}(X''Y_e^{c}\bar{\rho}_{c} +
2X'Y_e^{man}\bar{\rho}_{man}) - 
2\Delta E'X' - \Delta E''X'']$$ 
\vspace{-1pc}
$$~~~~~~~~~~~~~~
- {\Delta m^2_{21}\over{2E}}~X\cos 2\varphi_{12}~,~\eqno(9)$$
\noindent where $X = X'' + 2X'$ and 
$\cos 2\varphi_{12} = (|U_{e1}|^2 - |U_{e2}|^2)/(1 - |U_{e3}|^2)$.   
The interpretation of the Super-Kamiokande
data [6] in terms of $\nu_{\mu} \leftrightarrow \nu_{\tau}$
oscillations requires 
$s^2_{23} \cong (0.30 - 0.70)$, with 0.5 being the statistically 
preferred value. For the predicted
ratio $r(E,\theta_{z})$ of the 
$\nu_{\mu}$ and $\nu_e$ fluxes 
for $-1 \leq \cos\theta_{z}\ltap -0.8$ one has 
[19,20]: $r(E,\theta_{z}) \cong (2.0 - 2.5)$ for 
the neutrinos contributing to the sub-GeV event
samples, and $r(E,\theta_{z}) \cong (2.6 - 4.5)$ for those 
giving the main contribution to the multi-GeV samples.
If $s^2_{23} = 0.5$ and $r(E,\theta_{z}) \cong 2.0$
we have $(s^2_{23}~r(E,\theta_{z}) - 1) \cong 0$ and
the possible effects of the 
$\nu_{\mu} \rightarrow \nu_{e}$ 
and $\nu_{e} \rightarrow \nu_{\mu (\tau)}$ 
transitions on the $\nu_e$ flux and correspondingly on 
the sub-GeV $e-$like event sample, 
would be strongly suppressed even 
if these transitions are maximally
enhanced by the NOLR. 
The indicated suppression may actually be taking place
\footnote{Indeed, a more detailed investigation [18] 
performed within the indicated three-neutrino mixing scheme
reveals, in particular, that the excess 
of e-like events in the Super-Kamiokande sub-GeV data 
at $-1 \leq \cos\theta_{z}\leq -0.6$
seems unlikely to be due to small mixing angle,
$\sin^22\theta_{13} \ltap 0.20$, $\nu_{\mu} \rightarrow \nu_{e}$
transitions amplified by the oscillation length resonance.}.
Note, however, 
that the factor
$(s^2_{23}~r(E,\theta_{z}) - 1)$ can be as large as
$\sim 0.6$ for the sub-GeV neutrinos and that 
the effects of the NOLR enhanced
$\nu_{\mu (e)} \rightarrow \nu_{e (\mu;\tau)}$ 
transitions in the sub-GeV $e-$like event sample 
can be suppressed by the specific way this sample is 
selected from the data in 
the Super-Kamiokande experiment [18]. 
For the multi-GeV neutrinos we have 
$(s^2_{23}~r(E,\theta_{z}) - 1) \gtap 0.3~(0.9)$ for 
$s^2_{23} = 0.5~(0.7)$.  
The factor $(s^2_{23}~r(E,\theta_{z}) - 1)$ 
amplifies the effect of the 
$\nu_{\mu} \rightarrow \nu_{e}$ 
transitions in the $e-$like sample 
for $E \gtap (5 - 6)~{\rm GeV}$, for which
$r(E,\theta_{z}) \gtap 4$ [19,20]. 
This discussion suggests that 
the effects of the neutrino 
oscillation length resonance
in the Super-Kamiokande $e-$like multi-GeV data 
for $\Delta m^2_{31} \sim (2 - 6)\times10^{-3}~{\rm eV^2}$ 
should be much larger than those  
in the $e-$like sub-GeV data
for $\Delta m^2_{31} \sim (0.5 - 1.0)
\times 10^{-3}~{\rm eV^2}$ [2,3,18]. 
As similar analysis of the expression (8) for 
$\Phi_{\nu_{\mu}}$ shows,
they are also expected to be larger
(within the 3$-\nu$ mixing scheme considered)
than the NOLR 
effects in the
multi-GeV $\mu-$like event sample [2,3,18]. In all samples,
except, perhaps, the sub-GeV $e-$like one, the 
NOLR effects can be observable. 

\vspace*{-4pt}
\section{Conclusions}
\vspace*{-4pt}

  The neutrino oscillation length resonance should be
present in the $\nu_{2} \rightarrow \nu_{e}$
transitions 
of the solar neutrinos
crossing the Earth core on the way to the detector,
if the solar neutrino problem is due to 
small mixing angle MSW 
$\nu_{e} \rightarrow \nu_{\mu}$ 
transitions in the Sun.     
The same resonance should be operative also
in the $\nu_{\mu} \rightarrow \nu_{e}$
($\nu_{e} \rightarrow \nu_{\mu (\tau)}$) 
small mixing angle transitions 
of the atmospheric neutrinos crossing the Earth
core if the atmospheric $\nu_{\mu}$ and 
$\bar{\nu}_{\mu}$ indeed take part 
in large mixing vacuum 
$\nu_{\mu}(\bar{\nu}_{\mu}) \leftrightarrow \nu_{\tau}(\bar{\nu}_{\tau})$
oscillations with 
$\Delta m^2 \sim (0.5 - 6.0)\times 10^{-3}~{\rm eV^2}$,
as is strongly suggested by the Super-Kamiokande
data [6], and if all three flavour neutrinos are mixed in vacuum.
Actually, the resonance may have already
manifested itself in the excess of e-like events 
at $-1 \leq \cos\theta_{z}\leq -0.6$ observed
in the Super-Kamiokande multi-GeV atmospheric 
neutrino data [1,2,3,18]. 
And it can be responsible for at least part of the
strong zenith angle dependence present in the
Super-Kamiokande multi-GeV and sub-GeV $\mu-$like data [2,3,18]. 
\vspace*{-2pt}
\section{Acknowledgements.}
\vspace*{-6pt}
 The author would like to 
thank H. Minakata and O. Yasuda for  
their kind hospitality during  
the {\it New Era in Neutrino Physics} Symposium.  

\section{References}
\vspace*{-4pt}

\re
1. S.T. Petcov,
Phys. Lett. {\bf B434} (1998) 321. 
\re
2. S.T. Petcov, Talk given at the Workshop 
`` Neutrino Physics and Astrophysics: from Solar
to Ultra-High Energy'', Aspen Center for Physics,
June 29 - July 12, 1998, Aspen, U.S.A.
\re
3. M. Chizhov, M. Maris and S.T. Petcov, Report SISSA  53/98/EP, 31 July 1998
(hep-ph/9810501); see also: S.T. Petcov, 
hep-ph/9809587.
\re
4. M. Maris and S.T. Petcov, Phys. Rev. {\bf D56} (1997) 7444, 
and hep-ph/9803244.
\re
5. M. Nakahata (Super-Kamiokande Collaboration), these Proceedings.
\re
6. T. Kajita, (Super-Kamiokande Collaboration), these Proceedings.
\re
7. F.D. Stacey,
 {\it Physics of the Earth}, John Wiley and Sons, 
New York, 1977;
A.D. Dziewonski and D.L. Anderson,
  Phys. Earth Planet. Interiors
                {\bf 25} (1981) 297.
\re
8. P.I. Krastev and S.T. Petcov, Phys. Lett. 
{\bf B205} (1988) 84.
\re
9. M. Maris and S.T. Petcov, in preparation;
M. Maris, Q.Y. Liu and S.T. Petcov, 
study performed in December of 1996 (unpublished).
\re
10. R. Jeanloz,
Annu. Rev. Earth Planet. Sci. {\bf 18} (1990) 356;
C.J. All{\`e}gre et al., Earth
Planet. Sci. Lett. {\bf 134} (1995) 515.
\re 
11. S.T. Petcov, Lecture Notes in Physics, vol. {\bf 512}
(eds. H. Gausterer and C.B. Lang, Springer, 1998), p. 281.
\re
12. S.P. Mikheyev and A.Yu. Smirnov, Proc.  
of the Moriond Workshop on Massive Neutrinos, 1986 
(eds. O. Fackler and J. Tran Thanh Van, 
Editions Fronti{\`e}res, 
France, 1986), p. 355;
E.D. Carlson, Phys. Rev. {\bf D34} (1986) 1454;
A. Dar et al., Phys. Rev. {\bf D35} (1987) 3607;
M. Cribier et al.,
Phys. Lett. {\bf{B182}} (1986) 89;
A.J. Baltz and J. Weneser,
Phys. Rev. {\bf D35} (1987) 528; J.M. Gelb, W. Kwong 
and S.P. Rosen, Phys. Rev. Lett. {\bf 78} (1997) 2296.
\re
13. Q.Y. Liu, M. Maris and S.T. Petcov,
Phys. Rev. {\bf D56} (1997) 5991.
\re
14. V.K. Ermilova et al., Short Notices of 
the Lebedev Institute {\bf 5} (1986) 26; E.Kh. Akhmedov, Yad. Fiz. {\bf 47}
(1988) 475; P.I. Krastev and A.Yu. Smirnov, Phys. Lett. {\bf B226} (1989) 341.
\re
15. Q.Y. Liu and A.Yu. Smirnov, hep-ph/9712493.
\re
16. S.T. Petcov, Phys. Lett. {\bf B214} (1988) 259.
\re
17. M. Appolonio et al. (CHOOZ Collaboration), Phys. Lett.
{\bf B420} (1998) 397.
\re
18. S.T. Petcov, L. Wolfenstein and O. Yasuda, work in progress.
\re
19. M. Honda et al., Phys. Rev. {\bf D52} (1995) 4985.
\re 
20. V. Agraval et al., {\bf D53} (1996) 1314.

\end{document}